\documentclass{article}

\usepackage{arxiv}

\usepackage[utf8]{inputenc} 
\usepackage[T1]{fontenc}    
\usepackage{hyperref}       
\usepackage{url}            
\usepackage{booktabs}       
\usepackage{amsfonts}       
\usepackage{nicefrac}       
\usepackage{microtype}      
\usepackage{lipsum}		
\usepackage{graphicx}
\usepackage{natbib}
\usepackage{doi}

\title{Flat Teams Drive Scientific Innovation}

\author{ Fengli Xu \\
	Knowledge Lab, Department of Sociology\\
	University of Chicago\\
	Chicago, IL\\
	\texttt{fenglixu@uchicago.edu} \\
	\And
	Lingfei Wu \\
	School of Computing and Information\\
	University of Pittsburgh\\
	Pittsburgh, PA \\
	\texttt{liw105@pitt.edu} \\
	\AND
	James A. Evans \\
	Knowledge Lab, Department of Sociology\\
	University of Chicago\\
	Chicago, IL\\
	\texttt{jevans@uchicago.edu} \\
}

\renewcommand{\shorttitle}{Flat Teams Drive Scientific Innovation}

\hypersetup{
pdftitle={Flat Teams Drive Scientific Innovation},
pdfsubject={Science of Science},
pdfauthor={Fengli Xu, Lingfei Wu, James A. Evans},
pdfkeywords={teams, innovation, science of science, productivity, novelty, disruption},
}

\begin{document}
\maketitle

\begin{abstract}
With teams growing in all areas of scientific and scholarly research, we explore the relationship between team structure and the character of knowledge they produce. Drawing on 89,575 self-reports of team member research activity underlying scientific publications, we show how individual activities cohere into broad roles of (1) leadership through the direction and presentation of research and (2) support through data collection, analysis and discussion. The hidden hierarchy of a scientific team is characterized by its lead (or L)-ratio of members playing leadership roles to total team size. The L-ratio is validated through correlation with imputed contributions to the specific paper and to science as a whole, which we use to effectively extrapolate the L-ratio for 16,397,750 papers where roles are not explicit. We find that relative to flat, egalitarian teams, tall, hierarchical teams produce less novelty and more often develop existing ideas; increase productivity for those on top and decrease it for those beneath; increase short-term citations but decrease long-term influence. These effects hold within-person -- the same person on the same-sized team produces science much more likely to disruptively innovate if they work on a flat, high L-ratio team. These results suggest the critical role flat teams play for sustainable scientific advance and the training and advancement of scientists.  
\end{abstract}


\section{Introduction}
Teams are the engines of modern science, having grown in prevalence and size across all areas of scientific and scholarly investigation \cite{wuchty2007increasing}. Despite the known importance of team structure in many domains of economy and society, little is known about how team structure in science relates to innovation and discovery outcomes from a lack of consistent, large scale data. Previous experimental and observational studies of emergent team structure reveal that flatter teams with more balanced \cite{woolley2010evidence} or synchronous communication between members \cite{mayo2021variance,saavedra2011synchronicity} achieve higher performance in problem solving \cite{woolley2010evidence}, sales \cite{mayo2021variance}, trading \cite{saavedra2011synchronicity} and healthcare \cite{kreps1994effective} settings, partially because coordinated attention facilitates the adaptability needed to respond to uncertainty, complexity and change \cite{uitdewilligen2010team}. Hierarchy, by contrast, accelerates rapid top-down communication and efficiency \cite{anicich2015hierarchical}, but necessarily reduces symmetric coordination \cite{hackman1987design} and yields greater inequality for team member benefits, ranging from higher deaths in mountaineering expeditions \cite{anicich2015hierarchical} to uneven sacrifices in market contexts \cite{edge1984impact}. 

Calls for more transparent, honest and equitable credit from the open science movement have inspired increased mandatory reporting for individual research contribution on published papers in most high-profile journals. Reporting has become increasingly standardized to more accurately reflect researcher contribution and signal contributor skills. In this paper, we use contributor-level information to explore the relationship between the hierarchy of individual team contributions and the character of the team's contribution to unfolding scientific advance. Recent studies analyzed the division of labor across stated scientific contributions \cite{Lariviere2016-kr, Haeussler2020-ac}, but did not explore the hierarchical research roles that emerge from the inequality of contributions (e.g., ``core''/``lead'' versus ``extended''/``supporting'' team members \cite{milojevic2014principles, milojevic2018changing}).    

Here, we demonstrate how specific scientist contributions cohere into hierarchical roles that lead or follow in support of research publication and yield a simple lead (or L)-ratio associated with each paper of $n$ authors ranging from $1/n$ for maximum hierarchy to 1.0 for flatness. Teams with higher L-ratios broadly share leadership opportunities in fulsome collaboration, while those with low L-ratios segregate leading from supporting contributions. We validate these patterns with the position of authors in paper bylines, the imputed ideas and prior knowledge each scientist contributes to each paper, and the history of contributions scientists have made to science as a whole. These signals are available for all papers and enable robust extrapolation from papers with self-reports to all papers in science. These patterns reveal how team hierarchy may emerge as teams grow. When it does, contributions for science differ dramatically. 

Teams with higher L-ratios manifest greater novelty in the atypical combination of ideas \cite{Uzzi2013-wd}, while those with lower L-ratios engage in the development of established research directions \cite{Wu2019-al}. Teams with higher L-ratios facilitate greater productivity for the average author, while those with low L-ratios amplify the productivity of just those on top. Finally, teams with higher L-ratios are associated with the potential for long-term scientific influence, while those with lower L-ratios contribute to ensured short-term attention.

\section{Results}
\label{sec:results}

\begin{figure*}
\centering
\includegraphics[width=1\linewidth]{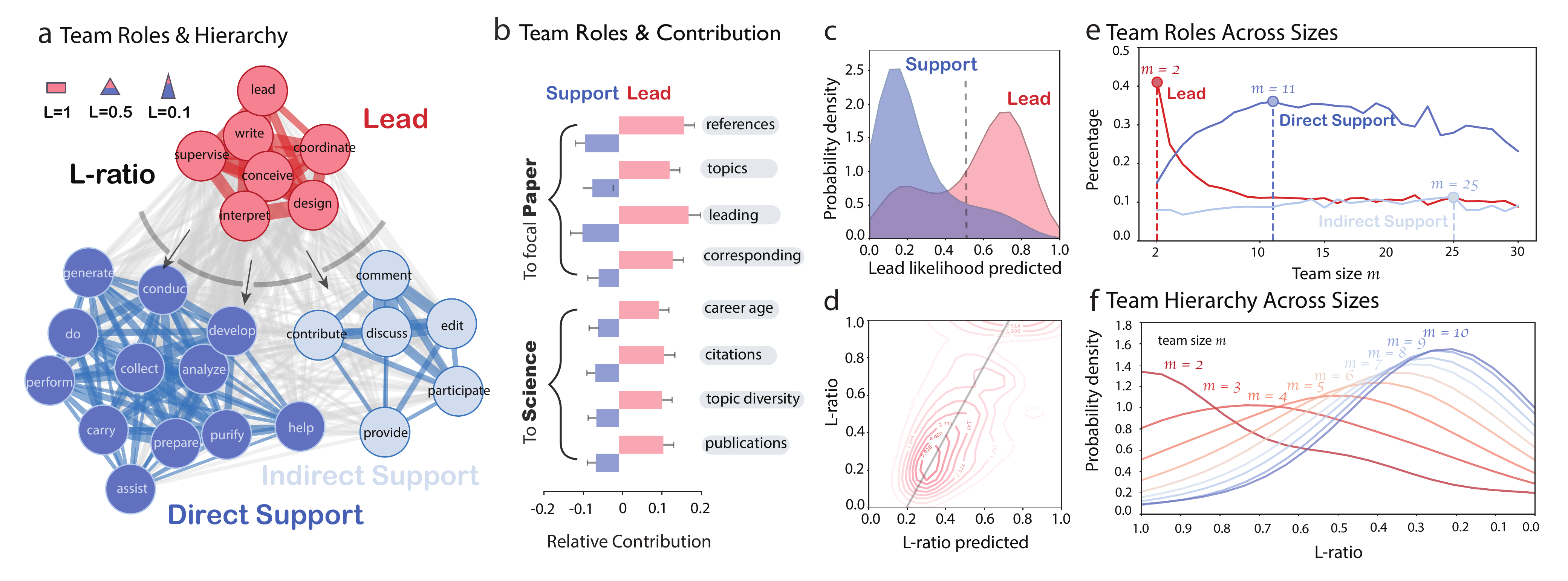}
\caption{\textbf{The hidden hierarchy of scientific teams.} (a) The co-occurrence of research activities within individual authors across 89,575 contribution statements. Three clusters including ``Lead (red),'' ``Direct Support (blue),'' and ``Indirect Support (light blue)'' are identified. Arrows imply the direction of influence. We verify L-ratio by demonstrating the distinct contributions of lead and support authors to specific papers and science as a whole (b). Our machine learning model classifies Lead and Support authors (precision 0.79, recall 0.793) and predicts L-ratio (Pearson correlation coefficient 0.66) (c-d). The composition of team roles (e) and the distribution of L-ratio (f) changes with team size.}
\label{fig:fig1}
\end{figure*}

Drawing on 89,575 self-reports of team member research activity underlying scientific papers published in \textit{PNAS}, \textit{Nature}, \textit{Science}, and \textit{PLOS One} from 2003 to 2020, we cluster the 25 most common research activities as a function of their co-contribution by authoring scientists. These activities cluster into broad roles of (1) leadership through the direction and presentation of research and (2) direct or indirect research support through data collection, analysis and discussion (see Fig. 1a). Specifically leadership involves the activities: ``conceive,'' ``design,'' ``lead,'' ``supervise,'' ``coordinate,'' ``interpret,'' and ``write.'' Direct and indirect support coherently separate into their own clusters. Direct support involves the activities: ``help,'' ``assist,'' ``prepare,'' ``develop,'' ``collect,'' ``generate,'' ``purify,'' ``carry,'' ``do,'' ``perform,'' ``conduct,'' and ``analyze.'' Indirect support activities occur before the research begins and after it is complete, including: ``participate,'' ``provide,'' ``contribute,'' ``comment,'' ``discuss,'' and ``edit.'' The cynical observer might reduce these roles to ``brain,'' ``muscle,'' and ``fat,'' the essential anatomy of modern research teams. By contrast, we demonstrate that when more members of the team are integrated into leading roles, the character of research changes and comes to influence the unfolding of scientific advance in strikingly different ways. The ``L-ratio'' quantifies the hierarchy of scientific teams, defined as the fraction of authors playing lead roles among all team members. 

Leading authors make measurably distinct contributions from those playing only supporting roles. Lead authors are 10\% to 20\% more likely than average to introduce references, direct topics, initiate research as first author, and manage communication as corresponding author. In contrast, support authors are 5\% to 10\% less likely than average to contribute to these tasks (Fig. 1b). We find a comparable distinction between lead and support roles when analyzing scientists' cumulative contribution to science measured in career age, citation impact, total number of topics studied, and total number of previously published papers. These characteristics allowed us to build a machine learning model to classify authors into ``Lead'' and ``Support'' roles (Fig. 1c), with precision of 0.79 and recall of 0.793, and to robustly predict the L-ratio of scientific teams (predicted and empirical L-ratios correlate at 0.66) (Fig. 1d). Using these models, we scale our measures of team L-ratios to the complete sample of 16,397,750 papers published during 1950-2020 where roles are not reported.

The composition of team roles changes with team size. The proportion of lead authors peak in teams of size two, authors exclusively playing direct support roles summit in teams of size 11, and those only in indirect support roles max at teams of 25 members (Fig. 1e). While L-ratio is clearly associated with team size such that smaller teams tend to have a higher L-ratio than larger teams (Fig. 1f), substantial variance in L-ratio for teams of the same size allows us to disentangle the relative effects of team hierarchy from size.       

\begin{figure*}
\centering
\includegraphics[width=1\linewidth]{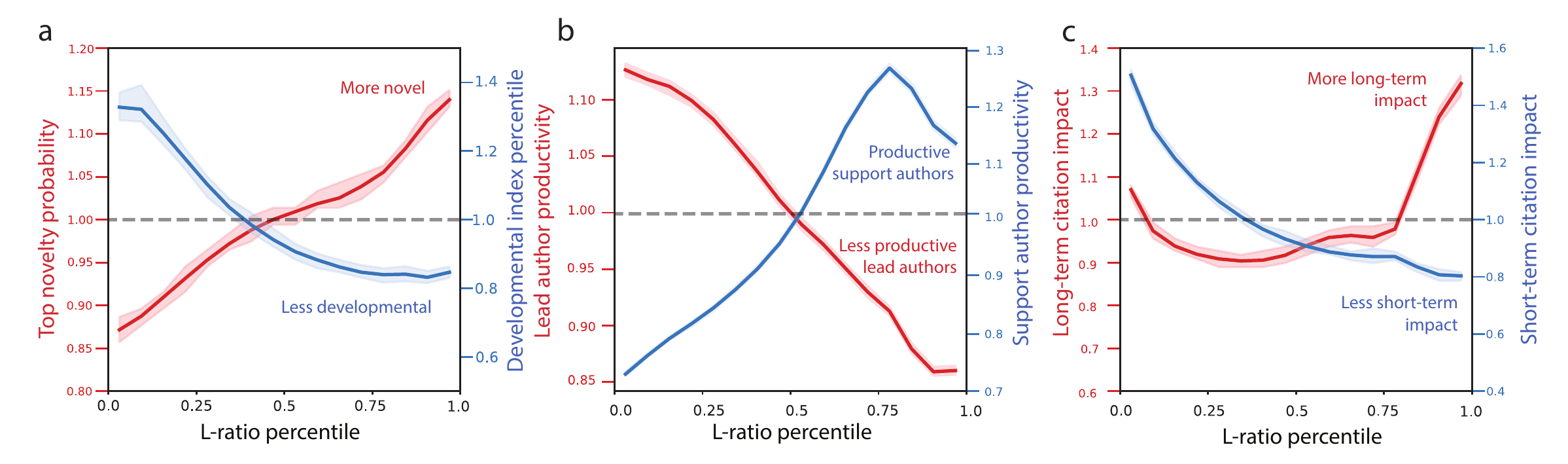}
\caption{\textbf{Tall vs. flat teams and the characters of research output.} Probability of writing a top 10\% novel paper (red) increases with L-ratio, whereas the percentile of development index (blue) decreases with it (A). Lead authors are less productive in teams with a higher L-ratio (red), whereas support authors experience productivity gains (blue) (B). Scientific publications from high L-ratio teams receive more long-term citations after twenty years (red) but fewer short-term citations within ten (blue) (C). Bootstrapped 95\% confidence intervals are shown as the shaded envelope for all curves.}
\label{fig:fig2}
\end{figure*}

Hierarchy is deeply related to characteristics of the resulting research and its recognition by others in science. The probability of writing a novel paper (top 10\% atypicality) increases with the team's L-ratio, while the likelihood that a team will be recognized by others as having incrementally developed rather than radically disrupted prior ideas, measured by the development index (the inverse of disruption score \cite{Wu2019-al}), decreases with it (Fig. 2a). We also find that lead authors are more productive in hierarchical teams with a lower L-ratio, but support authors experience greater productivity on flatter teams (Fig. 2b). Scientific publications from low L-ratio teams receive more short-term citations, while those from high L-ratio teams experience greater influence over the long term (Fig. 2c). We perform author fixed-effects regressions and find that the L-ratio continues to play a consistent and substantial role after controlling for team size and the career age variance of team members in predicting all outcome variables (see SI for details). This is important as changing the size or altering the age structure of a team for optimal innovative performance involves hiring or firing members, but our findings suggest the possibility that by simply re-organizing tasks, the same goal may be achieved: leading roles can be extended across large teams and involve younger scientists to maximize innovative potential.

\section{Discussion}
\label{sec:discussion}

Tall teams provide obvious benefits for scientists who lead them, but do not necessarily maximize the productivity and innovation potential of those who support. They enable greater lead author productivity, maximize immediate citations and so hedge against the risk that their work will not be credited. Under some circumstances, hierarchies may represent the most efficient allocation of effort for their research purposes, but they may impose costs on supporting scientists who do not play leadership roles and produce fewer papers. While the causality of these patterns remains unclear, our author fixed-effect models suggest that as the same researchers shift from teams with lower to higher L-ratios, their opportunities for leadership and productivity expand, corroborating prior research on the distribution of member risk associated with team hierarchy \cite{anicich2015hierarchical,edge1984impact}. 

Building on prior research about team member specialization \cite{Lariviere2016-kr, Haeussler2020-ac}, we uncover the hierarchy of research roles and compare the lasting contributions of tall versus flat teams. Tall teams produce less novel, more developmental, and shorter-lived contributions to science, suggesting that scaling teams to increase innovation poses a paradox, especially as sponsored science increasingly shifts from ``little'' to ``big'' in the name of accelerating advance. In addition to steady growth in team size \cite{wuchty2007increasing}, team hierarchy has also markedly increased over the past half century. If we define tall teams as those with an L-ratio below .5 -- where less than half of team members contribute to the conceptual work, these increase from 50\% in 1950 to 70\% in 2015. Concerns over scientific stagnation have arisen from apparent diminishing returns to scientific investment, inferred from accelerated growth in publications and aging scientists \cite{blau2017us} but slowed expansion in new ideas \cite{Chu2021-be}. Here we reveal the place of team hierarchy in the landscape of innovation, and the critical role flat teams play in advancing supporting scientists to grow next-generation scientific workforce for sustainable, long-term scientific advance.

\section{Methods}

\subsection{Datasets}
We link and analyze two data sets, Microsoft Academic Graph (MAG) and author contribution statements from four journals. Our MAG data include 16,397,750 journal articles written by two or more authors from 184 countries during 1950-2015. These papers contain 7.4 MAG topic keywords and 30.5 references on average, and the average number of citations to them is 41.2. The average team size (number of authors) is 4.8. We collect the author contribution statements from four journals, including \textit{PNAS}, \textit{Nature}, \textit{Science}, and \textit{PLOS ONE}. Our data cover a substantial body of all available contribution statements from the time they were required, covering 13 years of \textit{PNAS} (18,354 from 2003 to 2015), 15 years of \textit{Nature} (9,364 from 2006 to 2020), 3 years of \textit{Science} (1,176 from 2018 to 2020), and 9 years of \textit{PLOS ONE} (60,681 from 2006 to 2014). We apply NLP algorithms to extract the \emph{subject} and \emph{predicate} of each sentence and identify the research activities associated with authors. The top 25 activities (Fig. 1a) cover 94.6\% data and each paper lists 5.2 unique activities on average. The ``L-ratio,'' the fraction of Lead authors in the team, is 0.38 for Nature, 0.38 for Science, 0.45 for PNAS, and 0.43 for PLOS ONE.

\subsection{Predicting Lead vs. Support Roles and L-ratio}
We apply Modularity-based algorithm \cite{newman2006modularity} and identify three clusters from the co-occurred research activities, including ``Lead'', ``Direct Support'' and ``Indirect Support''. We analyze the distinct characteristics of Lead and Support authors (Fig. 1b) and build a neural network to predict author roles in 16,397,750 papers without explicit author contribution statements with precision of 0.79 and recall of 0.793. We extend this model by including team size and the unevenness of contributions and successfully predict L-ratio with a Pearson correlation coefficient equaling 0.66 between predicted and empirical values. 

\subsection{Quantifying Team Performances: Novelty, Development, Citations}

Novelty (Fig.2a) quantifies to what extent an paper draws upon and combines atypical ideas \cite{Uzzi2013-wd}. We applied a skip-gram word2vec model to learn the vector representation of MAG topic keywords from their co-occurrence within papers. The typicality of a pair of topic keywords is calculated as the inner product of their embedding vectors and novelty is defined as negative typicality. The developmental index (Fig.2a) is defined as the inverse of disruption \cite{Wu2019-al}. The Developmental index measures to what extend a paper incrementally refines previous work rather than radically challenging it. Productivity (Fig.2b) is measured as the total number of papers published by an author in the same year when they contributed to the specific paper under analysis. We separately analyzed the productivity of lead authors and support authors for each paper (Fig. 2b). We measure short-term impact as the number of citations a paper receives within ten years and long-term impact as citations received after twenty years. 

\subsection{Evaluating the Impact of L-ratio on Team Performances}

We evaluate the impact of L-ratio on different team performances by regressing novelty, developmental index, productivity of lead authors, productivity of support authors, short-term citation impact and long-term citation impact on L-ratio, respectively. For each model, we also control for team size, mean career age, and the standard deviation of career ages by including them together with L-ratio as independent variables. Finally, we perform fixed-effect regression to examine the impact of L-ratio for the same person moving between different teams. 

\section{Acknowledgment}
We are grateful for support from AFOSR (\#FA9550-19-1-0354), NSF (\#1829366 and \#1800956), and DARPA (\#HR00111820006). L. W. acknowledges the support of Pitt Cyber and Richard King Mellon Foundation.

\bibliographystyle{unsrtnat}
\bibliography{main}  






\end{document}